\documentclass[a4paper,11pt]{article}

\usepackage{pos}
\usepackage{siunitx}
\usepackage{caption}
\usepackage{subcaption}
\usepackage{sidecap}

\sidecaptionvpos{figure}{c}

\title{Measurement of the astrophysical diffuse neutrino flux in a combined fit of IceCube’s high energy neutrino data}

\ShortTitle{A combined fit of IceCube’s high energy neutrino data}

\author{The IceCube Collaboration \\{\normalsize \normalfont(a complete list of authors can be found at the end of the proceedings)}\\}

\emailAdd{rnaab@icecube.wisc.edu}
\emailAdd{erik.ganster@icecube.wisc.edu}
\emailAdd{zelong.zhang@icecube.wisc.edu}

\abstract{

The IceCube Neutrino Observatory has discovered a diffuse neutrino flux of astrophysical origin and measures its properties in various detection channels. With more than 10 years of data, we use multiple data samples from different detection channels for a combined fit of the diffuse astrophysical neutrino spectrum. This leverages the complementary information of different neutrino event signatures. For the first time, we use a coherent modelling of the signal and background, as well as the detector response and corresponding systematic uncertainties. The detector response is continuously varied during the simulation in order to generate a general purpose Monte Carlo set, which is central to our approach. We present a combined fit yielding a measurement of the diffuse astrophysical neutrino flux properties with unprecedented precision.
\vspace{4mm}
{\bfseries Corresponding authors:}
Richard Naab$^{1*}$, Erik Ganster$^{2}$, Zelong Zhang$^{3}$\\
{$^{1}$ \itshape DESY Zeuthen}\\
{$^{2}$ \itshape RWTH Aachen University}\\
{$^{3}$ \itshape Stony Brook University}\\[4mm]
$^*$ Presenter

\ConferenceLogo{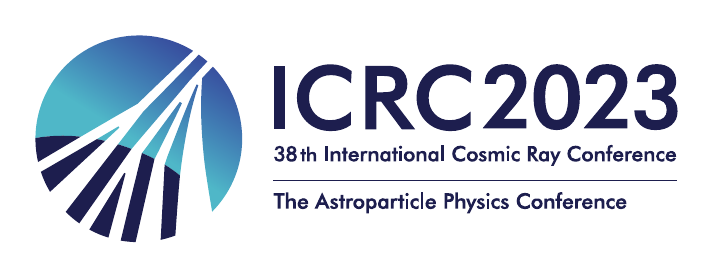}

\FullConference{The 38th International Cosmic Ray Conference (ICRC2023)\\ 26 July -- 3 August, 2023\\ Nagoya, Japan}
}

\begin{document}
\maketitle

\section{Introduction}\label{sec1}

A high-energy astrophysical neutrino flux was discovered by IceCube in 2013 \cite{Aartsen:2013jdh}, and has subsequently been characterized using several neutrino detection channels such as high energy starting events \cite{Abbasi:2020jmh}, through-going muon tracks \cite{IceCube:2021uhz} and cascades \cite{Aartsen:2020aqd}. The astrophysical neutrino flux has been well-described by an unbroken single power law (SPL) so far; see Figure \ref{fig:spl_overview} for an overview of the measured properties. Hints for substructure in the spectrum, as seen in independent measurements \cite{IceCube:2021uhz} and \cite{Aartsen:2020aqd}, are so far statistically insignificant, however.

Previously, several data samples have been combined in a likelihood fit \cite{Aartsen:2015knd}, thus enabling us to derive constraints from complementary data samples with different energy resolutions, sky coverage, and backgrounds. While this did result in an improved characterization of the astrophysical flux, not all uncertainties in the modeling were treated consistently: most notably, the energy scale was treated in an independent way between the samples, which increased the fit uncertainty significantly.

%MA: mention that it also is a lot more data than the 2015 fit 
Here, we present a combined fit using IceCube's track and cascade data, which for the first time uses a rigorously combined treatment of the detector response, as well as signal and background fluxes. The toolkit developed for such an analysis was previously verified, especially in terms of the treatment of detector systematic uncertainties \cite{GlobalFit_TeVPA22_erik, GlobalFit_TeVPA22_richard}, and can be easily extended to include more data samples in the future.

\section{Event samples}
\vspace*{-2mm}

The combined fit presented here uses events from the cascade and through-going track sample, with the event selections outlined in more detail below. By combining these two channels we leverage the complementary information from these different neutrino event signatures: shower-like events in the cascade channel provide good energy resolution because most of the visible energy is deposited within the instrumented volume of the detector.

Only atmospheric $\nu_\mu$ neutral current (NC) events, as well as all atmospheric $\nu_e$ events, produce shower-like events. For these cascade event signatures, atmospheric neutrinos can be vetoed effectively by observing accompanying muons from air showers; this so-called "self-veto" technique further reduces background for down-going events, see \cite{Arguelles:2018awr} and references therein. The track sample, on the other hand, provides a much larger effective area since $\nu_\mu$ interacting far outside the instrumented volume can produce high-energy muons which leave a track-like signature in the detector. The angular resolution for these events is much better and enables the resolution of the zenith-dependent atmospheric flux properties.

% TODO: Explain observables better in the following?

%This might be obvious, but is also different from previous GlobalFit
Both samples were processed using IceCube Pass-2 re-processed data using the latest detector calibrations for all track and cascade data used within this analysis \cite{Aartsen_2020}. This allows for fitting both selections with a common set of nuisance parameters modeling not only the atmospheric flux uncertainties but also the detector systematic uncertainties in a common and unified way.

\vspace*{-3mm}
\subsection{Track sample}
\vspace*{-2mm}
The through-going muon track sample described and defined in \cite{IceCube:2021uhz} focuses on up-going track-like events with a reconstructed zenith angle $\theta_{\mathrm{reco}} > \SI{85}{\degree}$. This cut uses the Earth as a shield against the background of atmospheric muons reaching the IceCube in-ice neutrino detector. This background is further reduced by a boosted decision tree (BDT) trained to separate atmospheric muons from muons originating from charged current muon-neutrino interactions. The result is a high purity (>\SI{99.8}{\percent}) sample of muon neutrinos of either atmospheric or astrophysical origin \cite{IceCube:2021uhz}. One year of data taken with the partial 59-string configuration of IceCube is not included in our combined fit since it cannot be described in the otherwise uniform Monte Carlo modeling of the detector response. In the combined fit, we use 542066 selected track events.
% This has, however, been shown to have negligible impact on the fit, as demonstrated in \ref{GlobalFit_TeVPA22}.

\vspace*{-3mm}
\subsection{Cascade sample}
\vspace*{-2mm}

We select cascade data as described in \cite{Aartsen:2020aqd}, resulting in a full-sky sample. The selection differentiates low-level cascade events in low- and high-energy regimes, where the low-energy event selection mainly uses a BDT method and the high energy ($E_\mathrm{reco} > \SI{60}{\tera\eV}$) event selection uses selection cuts on individual reconstructed event properties; see \cite{Aartsen:2020aqd} and references therein. The event selection was updated in terms of using the Pass-2 re-processing described above and uses improved modeling \cite{YUAN2023168440} of glacial ice for the reconstruction of cascade events, and the data sample was extended until mid 2021. In the combined fit, we use 12641 selected cascade events.

In addition to using events classified as cascades for spectral measurement, the low-energy event selection also identifies a class of events with muons mimicking showers. This muon control sample is used to constrain the flux level in our background modeling. Moreover, the single-sample cascade analysis also uses classified starting track events mimicking showers identified in the low-energy selection. This constrains the contribution of NC $\nu_\mu$ events constituting the dominant background from the conventional atmospheric neutrino flux. This latter class of events is not included in the combined fit presented here.

\section{Analysis methods}

\vspace*{-3mm}
\subsection{Detector response modeling and treatment of systematics}
\vspace*{-2mm}

In order to model the detector response and corresponding uncertainties, we use the so-called StormStorm MC approach, as outlined in \cite{Aartsen:2019jcj}. Detector response parameters which are uncertain are varied continuously on an event-by-event basis in the Monte Carlo simulation, populating the phase space in which the detector response parameters fall, which we constrain from calibration measurements. Contrary to the method proposed in \cite{Abbasi:2021n1}, we extract gradients from the SnowStorm simulation set \cite{Aartsen:2019jcj} and apply those to a separate Monte Carlo set simulated at nominal detector response parameters. This latter approach was verified in \cite{GlobalFit_TeVPA22_erik}. The detector response parameters taken into account in this analysis are those describing the bulk scattering and absorption coefficients of the glacial ice, the impact of the re-frozen drill holes as well as the efficiency of the optical sensors.

We use this Monte Carlo simulation consistently for estimating the observable distribution in all event selections, and we note that such an approach is central to the combined fit which has to take into account the correlations between the different event samples.

In order to simulate signal and background neutrino interaction events, we use the CSMS cross-section \cite{Cooper-Sarkar:2011jtt} and propagate secondary particles using \cite{Koehne:2013gpa}, and its parametrization of the muon energy losses.

\vspace*{-3mm}
\subsection{Background modelling}
\vspace*{-2mm}

Table \ref{tab:fit_parameters} lists all nuisance parameters used to model the background flux components in the combined fit.
The atmospheric neutrino flux contributions are calculated using MCEq \cite{Fedynitch:2015}. We use H4a as a primary cosmic ray flux model and Sibyll 2.3c as a hadronic interaction model as baseline, as well as an interpolation between the two primary cosmic ray models H4a and GST4 as in \cite{IceCube:2021uhz}; see also the references therein.
Uncertainties on the conventional neutrino flux are taken into account following the scheme from Barr et al. \cite{barr2006}, the corresponding nuisance parameters referred to as $\mathrm{Barr}\, X$, see \cite{IceCube:2021uhz} for detailed description of this as well as the muon template.

Modeling the effect of the atmospheric self-veto for the cascade selection is computationally challenging since, in principle, full air shower simulation including neutrino yields has to be produced to evaluate the effect of accompanying muons to atmospheric neutrinos. Since the event selections are typically very effective in terms of rejecting atmospheric muons, this becomes a hard problem, since most of the generated simulation will not pass the selection. In order to model the effect of the atmospheric self-veto, we use the MCEq atmospheric flux and muon range calculations employed in \cite{Arguelles:2018awr}.

We note that the response to atmospheric muons is event selection specific. It is parameterized in terms of the veto probability as a function of the muon energy at detector entry. The self-veto effective threshold describing this response is hard to constrain due to the computational limitations explained above; we therefore leave it a free nuisance parameter in the fit.
% Self-veto effect modelling expand method of continuously varying threshold, this selection specific response to muons is the most important parameter in the modelling and thus allow freedom due to computational limitations for determining the self-veto effect more precisely

\begin{table}
    \centering
    {\small
    \begin{tabular}{lccc}
        Nuisance parameter & Allowed Range & Prior \\
        \hline
        Optical Efficiency & $[0.9, 1.1]$ & -\\
        Bulk Ice Absorption & $[0.9, 1.1]$ & -\\
        Bulk Ice Scattering & $[0.9, 1.1]$ & -\\
        Hole-Ice $p_0$ & $[-0.84, +0.3]$ & -\\
        Hole-Ice $p_1$ & $[-0.134, +0.05]$ & -\\
        Self-veto Effective Threshold in GeV & $[5, 2000]$ & -\\
        Conventional Flux Normalization & $[0.0, \infty)$ & -\\
        Prompt Flux Normalization & $[0.0, \infty)$ & - \\
        Muon Flux Normalization (cascades only) & $[0.0, \infty)$ & -\\
        Muon Template Normalization (tracks only) & $[0.0, \infty)$ & -\\
        Cosmic-Ray Model Interpolation & $[-1.0, +2.0]$ & $\mathcal{G}(0.0, 1.0)$\\ 
        Cosmic-Ray Spectral Index Shift & $[-1.0, +1.0]$ & -\\
        Barr H & $[-0.8, +0.8]$ & $\mathcal{G}(0.0, 0.15)$\\
        Barr W & $[-0.6, +0.6]$ & $\mathcal{G}(0.0, 0.4)$\\
        Barr Y & $[-0.6, +0.6]$ & $\mathcal{G}(0.0, 0.3)$\\
        Barr Z & $[-0.244, +0.6]$ & $\mathcal{G}(0.0, 0.12)$\\
        \hline
    \end{tabular}
    }%small
    \caption{Nuisance parameters in the analysis, their allowed ranges, and Gaussian priors $\mathcal{G}(\mu, \sigma)$ (if used).}
    \label{tab:fit_parameters}
\end{table}

\vspace*{-3mm}
\subsection{Signal modelling}
\label{sec:astro_flux_models}
\vspace*{-2mm}

For modeling the astrophysical neutrino flux multiple models of the spectral shape are tested. For all models, the astrophysical flux is assumed to be isotropic over the full sky with a flavor composition of $(\nu_e:\nu_\mu:\nu_\tau)_\mathrm{at \, Earth} = (1:1:1 )$ and equal ratio of $\nu/\bar{\nu}$ corresponding to idealized pp sources where neutrinos are produced from decays of $\pi^{\pm}$.

The tested spectral shapes include a single power-law (SPL) flux, a log-parabolic (LogP), and a broken power law (BPL) flux according to models A, C, and D in \cite{Aartsen:2020aqd}. Also, a segmented flux model that assumes an $E^{-2}$ flux in individual neutrino energy bins with independent normalization each is used.

\vspace*{-3mm}
\subsection{Likelihood fit}
\vspace*{-2mm}

The analysis is performed with a forward-folding likelihood fit where the Monte Carlo events are weighted according to the parameters describing the signal and background fluxes as described above. We bin cascade and track data separately in the respective reconstructed energy and zenith observables, in order to resolve spectral features and discriminate the isotropic astrophysical neutrino flux from the atmospheric backgrounds with different zenith dependencies. We take into account the limited amount of Monte Carlo simulation statistics, due to computational limitations, in terms of using the effective likelihood described in \cite{Arguelles:2019izp}.
The analysis bins in the combined fit are  de-correlated by removing all overlapping events from the tracks sample - these are a minor fraction of starting events that are not relevant to the sensitivity of this measurement \cite{Abbasi:2021n1}. We note that this de-correlation can only be achieved using the general-purpose MC simulation developed for the purpose of this work.

% This is kinda optional
In order to perform the fit, $\mathcal{O}(10^7)$ Monte Carlo events have to be re-weighted at every minimizer iteration. In order to do this computationally efficiently, we use \textit{aesara} \cite{WillardAesara2023} allowing for fast evaluation of mathematical expressions involving multi-dimensional arrays as well as efficient symbolic differentiation, which we use to provide gradients of the likelihood function to the minimizer.
% TODO: also cite LBFGSB minimization algorithm here...?

\section{Results}

\begin{figure}
    \centering
    \begin{subfigure}[b]{0.48\textwidth}
        \centering
        \includegraphics[width=\textwidth]{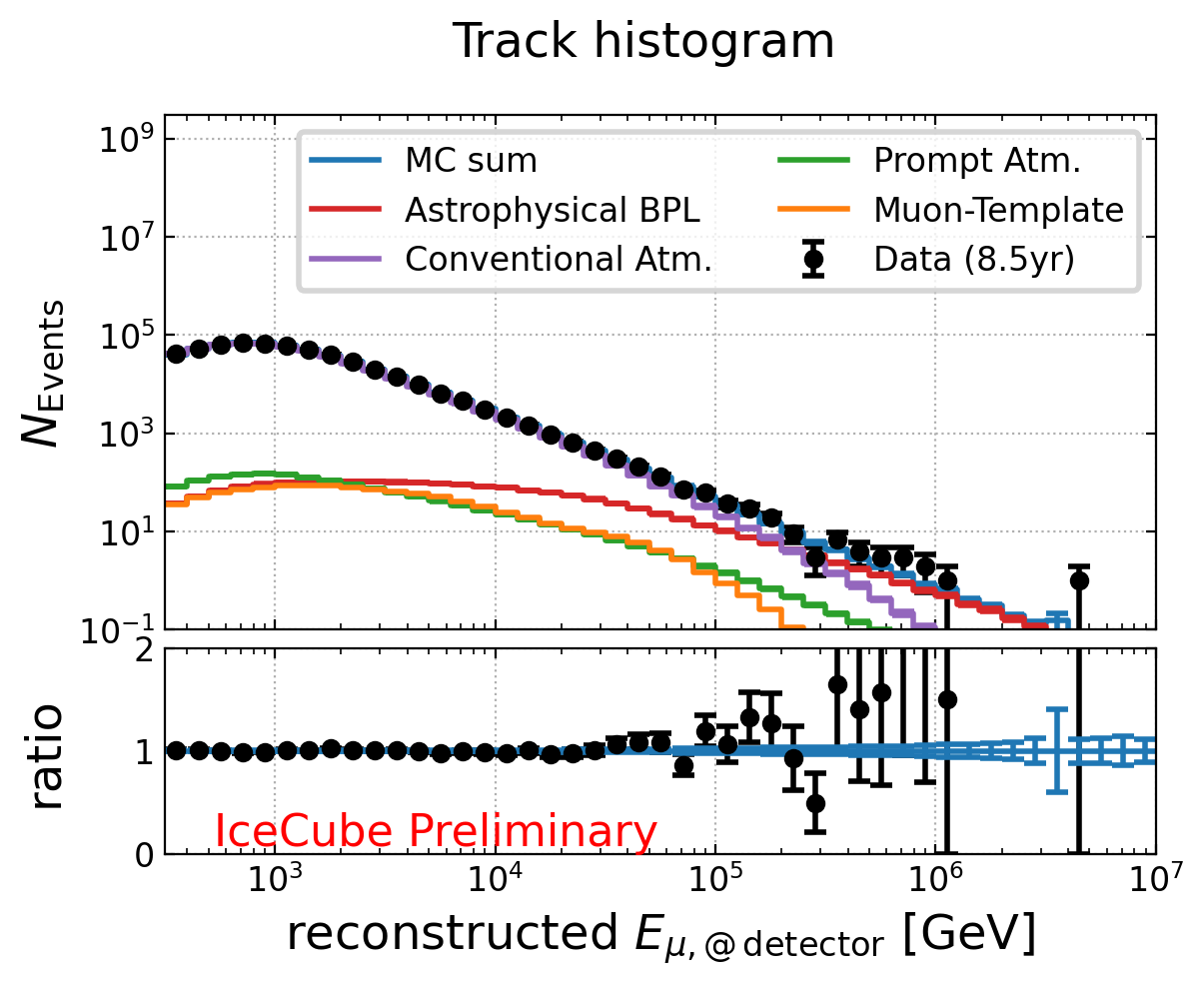}
        % \caption{Best-fit reconstructed muon energy spectrum for all events in the tracks sample assuming a broken power law astrophysical neutrino flux.}
        % \label{fig:tracks_spectrum_BPL}
    \end{subfigure}
    \hfill
    \begin{subfigure}[b]{0.48\textwidth}
        \centering
        \includegraphics[width=\textwidth]{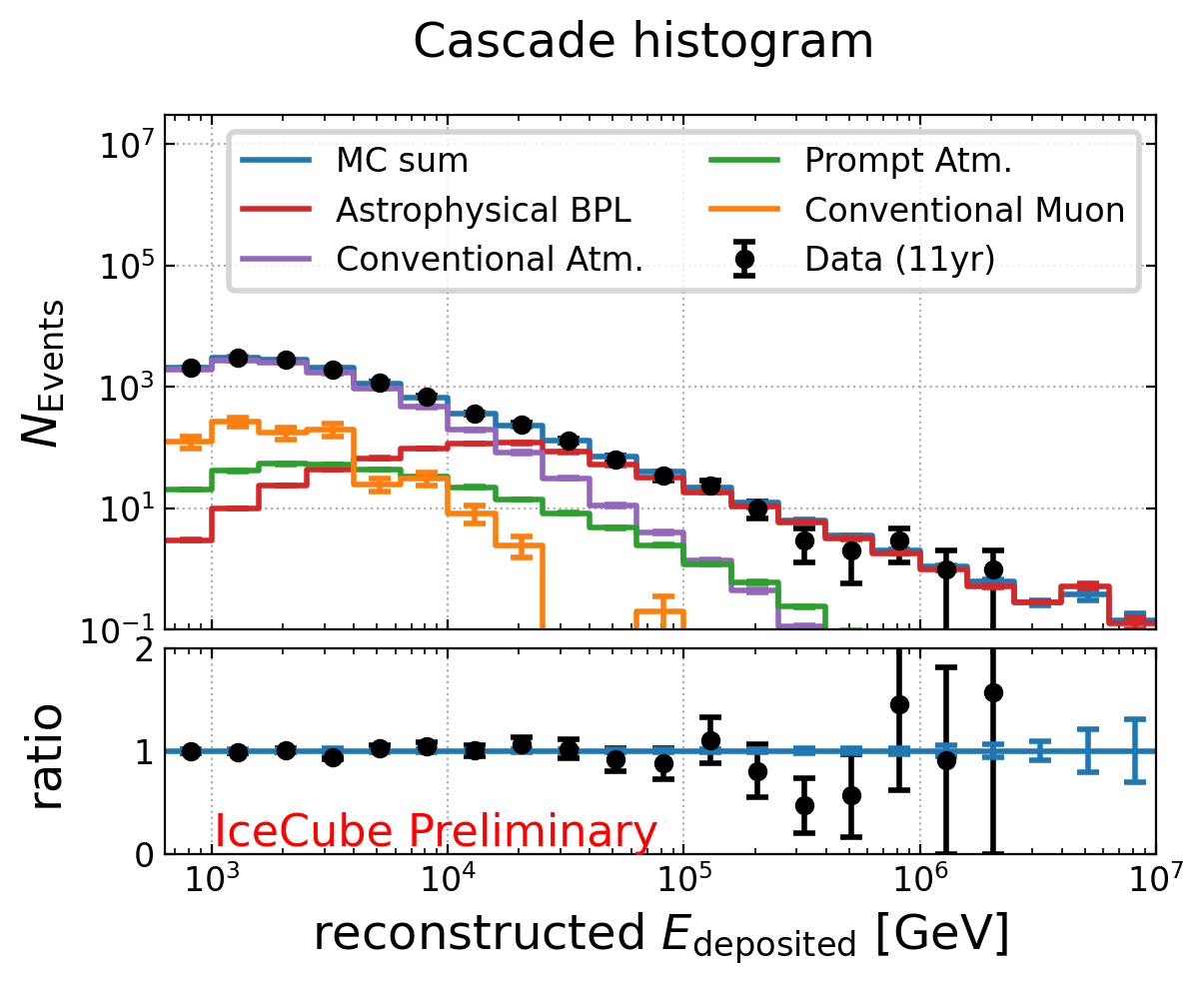}
        % \caption{Best-fit reconstructed deposited energy spectrum for all events in the cascade sample assuming a broken power law astrophysical neutrino flux.}
        % \label{fig:cascades_spectrum_BPL}
    \end{subfigure}
    % \vspace*{-5mm}
    \caption{Best-fit spectra assuming a BPL astrophysical neutrino flux. Left: Reconstructed muon energy spectrum for all events in the tracks sample. Right: Reconstructed deposited energy spectrum for all events in the cascade sample.}
    \label{fig:spectrum_BPL}
\end{figure}

We find good agreement between data and Monte Carlo in the combined fit, see Figure \ref{fig:spectrum_BPL}. This agreement was tested extensively in the background region below 10 TeV in reconstructed energy to ensure correct combined modeling of the region where the atmospheric fluxes dominate.

In the region of the spectrum where the astrophysical flux dominates, we see structures deviating from a SPL. Notably, we see an excess of data at around 20-30 TeV in reconstructed cascade energy, which is consistent with previous findings \cite{IceCube:2014rwe, Aartsen:2020aqd}. Moreover, we see a deficit in the reconstructed cascade energy spectrum at a few hundred TeV, which was also reported in \cite{Abbasi:2020jmh}. The events in the tracks sample do not provide the energy resolution necessary to resolve these fine features but help in the combined fit by constraining the atmospheric flux and detector nuisance parameters due to the high statistics of the sample.

\begin{SCfigure}
% \begin{figure}[htbp]
    \centering
    \includegraphics[width=0.65\textwidth]{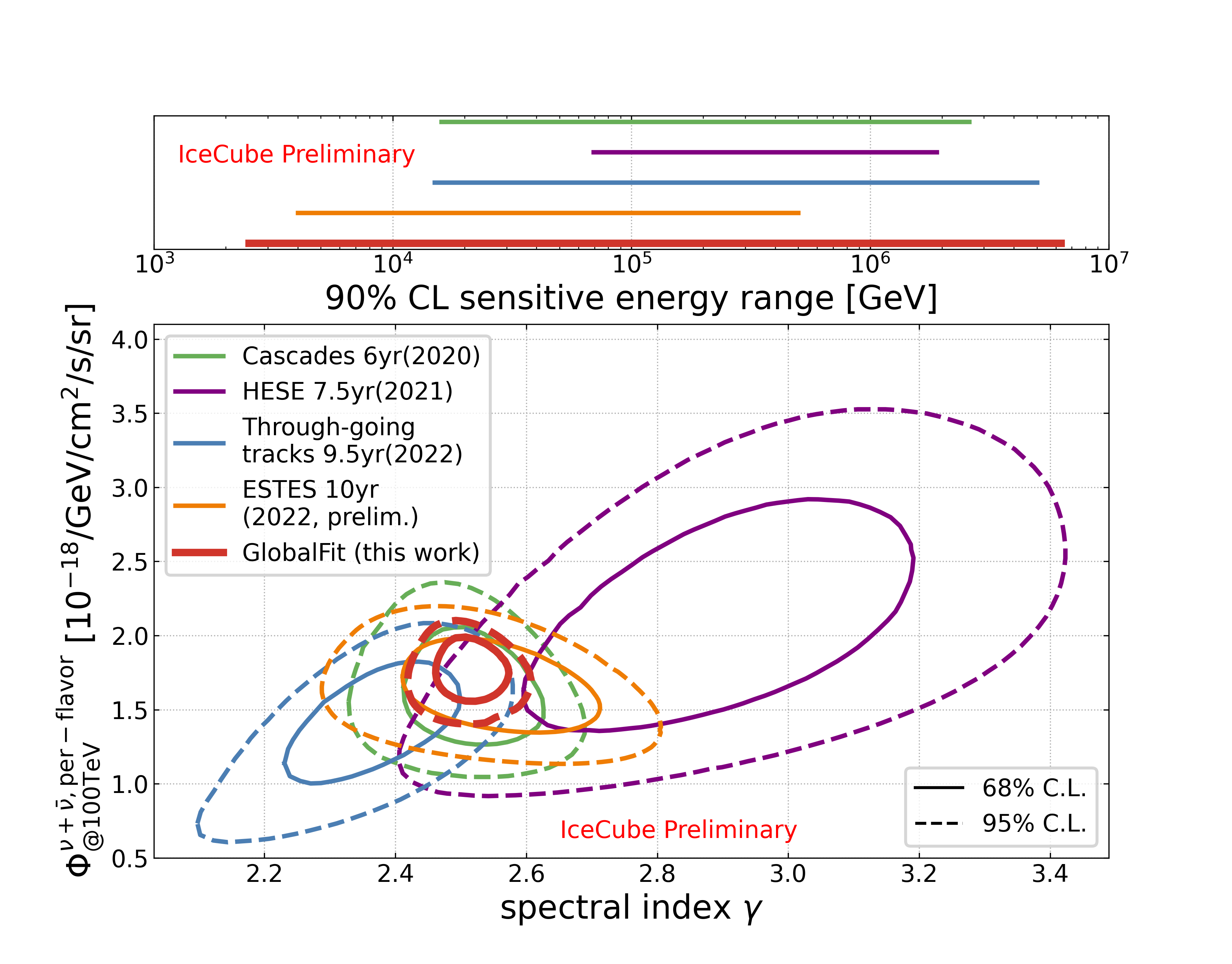}
    \caption{Result of the combined fit of tracks and cascades (in red) under the assumption of an astrophysical SPL neutrino flux. Previous results from measurements using single event channels are shown for comparison. Note that the sensitive neutrino energy ranges (as indicated in the upper panel) and neutrino flavor probed are different amongst the different samples.}
    % TODO: align notation with parametrisation in text
    \label{fig:spl_overview}
% \end{figure}
\end{SCfigure}

\begin{SCfigure}
% \begin{figure}[htbp]
    \centering
    \includegraphics[width=0.7\textwidth]{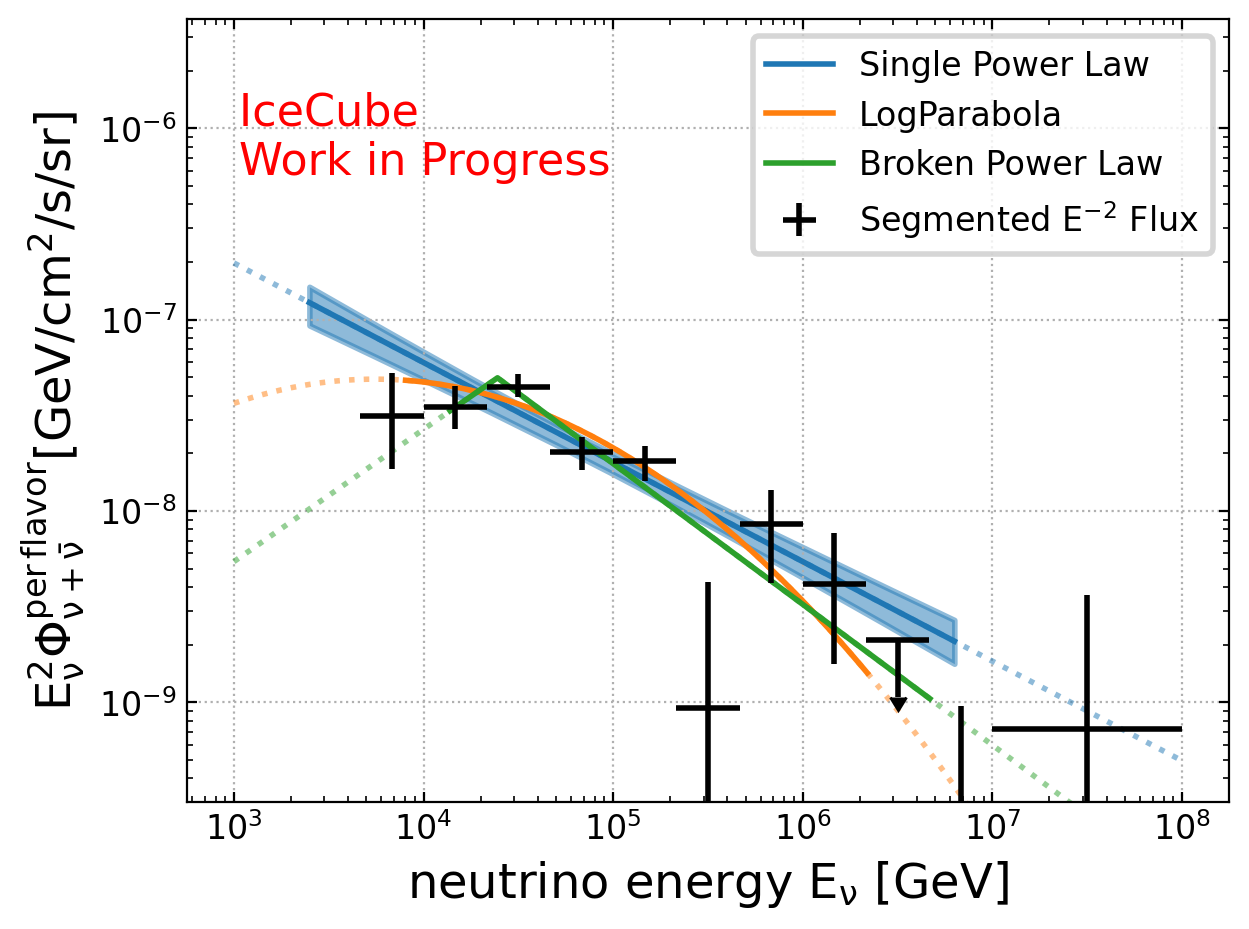}
    \caption{Result of the combined fit of tracks and cascades under different assumptions of the astrophysical neutrino flux. Solid lines represent the sensitive energy ranges of the corresponding astrophysical flux models. The uncertainty band shown in blue represents the 68\% CL uncertainties on the SPL fit. The segmented flux fit uncertainties are obtained by profiling single-segment normalizations over all other parameters in the fit.}
    \label{fig:astro_fluxes_true_energy}
% \end{figure}
\end{SCfigure}

The measured parameters of the SPL astrophysical neutrino flux are shown in Figure \ref{fig:spl_overview}. We note that more complex models are preferred statistically, but the contour shown here is in agreement with the sensitivity assuming a SPL. We test models of the astrophysical neutrino flux as described in Section \ref{sec:astro_flux_models}, and find that curvature or a spectral break in the astrophysical neutrino flux better describes the data. The outcome of these fits is described in Table \ref{tab:results} and also shown in Figure \ref{fig:astro_fluxes_true_energy}. The segmented flux fit follows this trend and indicates spectral features around 30 and several hundred TeV, as seen in the cascade energy spectrum. The sensitive energy ranges quoted in Table \ref{tab:results} are computed under the assumption of the respective fluxes. We fit the Asimov sets corresponding to astrophysical fluxes unrestricted in energy with the same spectral assumption, but restricted to $[E_{\nu,\mathrm{low}}, E_{\nu,\mathrm{up}}]$. The limits given in Table \ref{tab:results} are then derived from 1D profile likelihood scans of $E_{\nu,\mathrm{low}}$ and $E_{\nu,\mathrm{up}}$, assuming Wilk's theorem.

% Assuming a single power-law astrophysical neutrino flux, Eq. \ref{eq:spl}, we measure $\Phi_\mathrm{@100\mathrm{TeV}}^{\nu+\bar{\nu}}=1.80^{+0.13}_{-0.16}$ and $\gamma=2.52^{+0.04}_{-0.04}$. The sensitive energy range of our analysis to such an astrophysical neutrino flux ranges from 2.5 TeV to 6.3 PeV at 90\% CL.

% Assuming a log-parabolic astrophysical neutrino flux, Eq. \ref{eq:logp}, we measure $\Phi_\mathrm{@100\mathrm{TeV}}^{\nu+\bar{\nu}}=2.13^{+0.16}_{-0.19}$, $\alpha_\mathrm{LP}=2.57^{+0.06}_{-0.05}$ and $\beta_\mathrm{LP}=0.23^{+0.10}_{-0.07}$. The sensitive energy range of our analysis to such an astrophysical neutrino flux ranges from 8.0 TeV to 2.2 PeV at 90\% CL.

% Assuming a BPL astrophysical neutrino flux, Eq. \ref{eq:bpl}, we measure $\Phi_\mathrm{@100\mathrm{TeV}}^{\nu+\bar{\nu}}=1.77^{+0.15}_{-0.11}$, $\mathrm{log}_{10}(E_\mathrm{break} / \mathrm{GeV})=4.39^{+0.09}_{-0.08}$, $\gamma_1=1.31^{+0.50}_{-1.21}$ and $\gamma_2=2.74^{+0.06}_{-0.07}$. The sensitive energy range of our analysis to such an astrophysical neutrino flux ranges from 13.7 TeV to 4.7 PeV at 90\% CL.

The prompt neutrino flux component fits to a value close to the flux level expected from the decay of charmed mesons, but has a large uncertainty and also depends on the choice of the astrophysical flux model \cite{boettcher:2023icrc} since the prompt flux component is subdominant over the full energy range.

% All the uncertainties quoted above are derived from 1D profile likelihood scans, assuming Wilk's theorem.

\begin{table}[htbp]
\centering
{ \small 
    \begin{tabular}{c|c|c|c}
        \begin{tabular}{@{}c@{}}Astrophysical \\ Model\end{tabular} & Result & \begin{tabular}{@{}c@{}}Energy Range \\ (90\% CL)\end{tabular} & \begin{tabular}{@{}c@{}}$-2 \Delta \mathrm{log} \mathcal{L}$ \\ over SPL\end{tabular} \\
        \hline
        SPL & $\begin{array}{ccl}
        \Phi_\mathrm{@100\mathrm{TeV}}^{\nu+\bar{\nu}} \, / \,C \, &=&1.80^{+0.13}_{-0.16} \\
        \gamma&=&2.52^{+0.04}_{-0.04}
        \end{array}$ & 2.5 TeV to 6.3 PeV & - \\
        \hline
        LogP & $\begin{array}{ccl}
        \Phi_\mathrm{@100\mathrm{TeV}}^{\nu+\bar{\nu}} \, / \,C \, &=&2.13^{+0.16}_{-0.19} \\
        \alpha_\mathrm{LP}&=&2.57^{+0.06}_{-0.05} \\
        \beta_\mathrm{LP}&=&0.23^{+0.10}_{-0.07}
        \end{array}$ & 8.0 TeV to 2.2 PeV & 16.4 \\
        \hline
        BPL & $\begin{array}{ccl} 
        \Phi_\mathrm{@100\mathrm{TeV}}^{\nu+\bar{\nu}} \, / \,C \, &=&1.77^{+0.15}_{-0.11} \\
        \mathrm{log}_{10}(E_\mathrm{break} / \mathrm{GeV})&=&4.39^{+0.09}_{-0.08} \\
        \gamma_1&=&1.31^{+0.50}_{-1.21} \\
        \gamma_2&=&2.74^{+0.06}_{-0.07}
        \end{array}$ & 13.7 TeV to 4.7 PeV & 24.7 \\
    \end{tabular}
    } %small
    \caption{Measurement results for all tested astrophysical flux models and their estimated sensitive energy range. The uncertainties are derived from 1D profile likelihood scans, assuming Wilks' theorem. The flux normalization is measured in units of $C=\rm{10^{-18}/GeV/cm^2/s/sr}$.}
    \label{tab:results}
\end{table}

We note that the BPL assumption results in a description of the data that is nearly as good as the segmented flux differing only in $\sim$6 TS units, despite having fewer parameters. Fitting the samples individually, we find that a BPL fit to the cascade sample alone results in very similar astrophysical flux parameters, with a slightly softer $\gamma_2$. A BPL fit to only the tracks sample results in a higher $E_\mathrm{break}$, and also a slightly softer $\gamma_2$ compared to the combined fit. Performing a likelihood ratio test against the combined best-fit BPL, we see a $\sim$2 sigma difference in the BPL fit to only the tracks sample. These differences are also reflected in the energy spectra shown in Figure \ref{fig:spectrum_BPL}, where we see some excess of high-energy track events compared to the combined model.
We observe a mild tension between the low-energy spectral index $\gamma_1$, derived from cascades and tracks, and results derived from a sample of starting events (ESTES, \cite{estes:2023icrc}, this conference), where no significant break in the spectrum is observed. An important difference between these samples targeting different event signatures is that they reach the highest sensitivity in different regions of the sky. Possible future directions for investigations are, whether these differences can be attributed to non-isotropic contributions, such as the Galactic Plane \cite{Abbasi:2023bvn}, and/or to the uncertainties of modeling the atmospheric neutrino background, the neutrino inelasticity, muon energy loss modeling, as well as the optical properties of the glacial ice \cite{tc-2022-174}.

\section{Conclusion and Outlook}

For the first time, we perform a fit of two independent data samples of IceCube's high-energy neutrino data using a consistent treatment of nuisance parameters. We find good agreement between the data and our model, with some deviations at the highest energies. We see indications for deviations from a SPL with hardening at lower energies and softening at higher energies, in agreement with what has been reported before using the tracks and cascades samples going into our combined fit. This opens up possibilities for deepening our understanding of the high-energy neutrino flux. The toolkit compiled for this work will be expanded with more event selections in the future for even better constraints on the astrophysical neutrino flux properties.

% Bibtex references:
\bibliographystyle{ICRC}
\bibliography{references}

% Alternatively, you can include references by hand:
%\begin{thebibliography}{99}
%\bibitem{...}
%https://www.overleaf.com/project/6482ed524e8665d2f0fe4803
%\end{thebibliography}

\clearpage

%The following list of authors, affiliations and funding agencies will be updated at the day of submission. The following template is a placeholder generated via https://authorlist.icecube.wisc.edu/icecube on March 25, 2023 and will be updated.
\section*{Full Author List: IceCube Collaboration}

\scriptsize
\noindent
R. Abbasi$^{17}$,
M. Ackermann$^{63}$,
J. Adams$^{18}$,
S. K. Agarwalla$^{40,\: 64}$,
J. A. Aguilar$^{12}$,
M. Ahlers$^{22}$,
J.M. Alameddine$^{23}$,
N. M. Amin$^{44}$,
K. Andeen$^{42}$,
G. Anton$^{26}$
% ...
\\
\\
% ...
$^{12}$ Universit{\'e} Libre de Bruxelles, Science Faculty CP230, B-1050 Brussels, Belgium \\
% ...
$^{17}$ Department of Physics, Loyola University Chicago, Chicago, IL 60660, USA \\
$^{18}$ Dept. of Physics and Astronomy, University of Canterbury, Private Bag 4800, Christchurch, New Zealand \\
% ...
$^{22}$ Niels Bohr Institute, University of Copenhagen, DK-2100 Copenhagen, Denmark \\
$^{23}$ Dept. of Physics, TU Dortmund University, D-44221 Dortmund, Germany \\
% ...
$^{26}$ Erlangen Centre for Astroparticle Physics, Friedrich-Alexander-Universit{\"a}t Erlangen-N{\"u}rnberg, D-91058 Erlangen, Germany \\
% ...
$^{40}$ Dept. of Physics and Wisconsin IceCube Particle Astrophysics Center, University of Wisconsin{\textendash}Madison, Madison, WI 53706, USA \\
% ...
$^{42}$ Department of Physics, Marquette University, Milwaukee, WI, 53201, USA \\
% ...
$^{44}$ Bartol Research Institute and Dept. of Physics and Astronomy, University of Delaware, Newark, DE 19716, USA \\
% ...
$^{63}$ Deutsches Elektronen-Synchrotron DESY, Platanenallee 6, 15738 Zeuthen, Germany  \\
$^{64}$ Institute of Physics, Sachivalaya Marg, Sainik School Post, Bhubaneswar 751005, India \\
% ...

\subsection*{Acknowledgements}

\noindent
The authors gratefully acknowledge the support from the following agencies and institutions:
USA {\textendash} U.S. National Science Foundation-Office of Polar Programs,
U.S. National Science Foundation-Physics Division,
U.S. National Science Foundation-EPSCoR,
Wisconsin Alumni Research Foundation,
Center for High Throughput Computing (CHTC) at the University of Wisconsin{\textendash}Madison,
Open Science Grid (OSG),
Advanced Cyberinfrastructure Coordination Ecosystem: Services {\&} Support (ACCESS),
Frontera computing project at the Texas Advanced Computing Center,
U.S. Department of Energy-National Energy Research Scientific Computing Center,
Particle astrophysics research computing center at the University of Maryland,
Institute for Cyber-Enabled Research at Michigan State University,
and Astroparticle physics computational facility at Marquette University;
Belgium {\textendash} Funds for Scientific Research (FRS-FNRS and FWO),
FWO Odysseus and Big Science programmes,
and Belgian Federal Science Policy Office (Belspo);
Germany {\textendash} Bundesministerium f{\"u}r Bildung und Forschung (BMBF),
Deutsche Forschungsgemeinschaft (DFG),
Helmholtz Alliance for Astroparticle Physics (HAP),
Initiative and Networking Fund of the Helmholtz Association,
Deutsches Elektronen Synchrotron (DESY),
and High Performance Computing cluster of the RWTH Aachen;
Sweden {\textendash} Swedish Research Council,
Swedish Polar Research Secretariat,
Swedish National Infrastructure for Computing (SNIC),
and Knut and Alice Wallenberg Foundation;
European Union {\textendash} EGI Advanced Computing for research;
Australia {\textendash} Australian Research Council;
Canada {\textendash} Natural Sciences and Engineering Research Council of Canada,
Calcul Qu{\'e}bec, Compute Ontario, Canada Foundation for Innovation, WestGrid, and Compute Canada;
Denmark {\textendash} Villum Fonden, Carlsberg Foundation, and European Commission;
New Zealand {\textendash} Marsden Fund;
Japan {\textendash} Japan Society for Promotion of Science (JSPS)
and Institute for Global Prominent Research (IGPR) of Chiba University;
Korea {\textendash} National Research Foundation of Korea (NRF);
Switzerland {\textendash} Swiss National Science Foundation (SNSF);
United Kingdom {\textendash} Department of Physics, University of Oxford.

\end{document}